\documentclass[aps,prl,twocolumn,showpacs,superscriptaddress]{revtex4}
\usepackage{graphicx}
\usepackage{dcolumn}
\usepackage{bm}

\newcommand{\ket}[1]{| #1 \rangle}

\newcommand{\be}{\begin{equation}}
\newcommand{\ee}{\end{equation}}
\newcommand{\bea}{\begin{eqnarray}}
\newcommand{\eea}{\end{eqnarray}}
\newcommand{\beas}{\begin{eqnarray*}}
\newcommand{\eeas}{\end{eqnarray*}}
\newcommand{\bes}{\begin{equation*}}
\newcommand{\ees}{\end{equation*}}

\begin{document}

\title{Ultrafast switching of photonic entanglement}

\author{Matthew A. Hall}
\affiliation{Center for Photonic Communication and Computing, EECS Department\\
Northwestern University, 2145 Sheridan Road, Evanston, IL
60208-3118}
\author{Joseph B. Altepeter}
\affiliation{Center for Photonic Communication and Computing, EECS Department\\
Northwestern University, 2145 Sheridan Road, Evanston, IL
60208-3118}
\author{Prem Kumar}
\affiliation{Center for Photonic Communication and Computing, EECS Department\\
Northwestern University, 2145 Sheridan Road, Evanston, IL
60208-3118}

\begin{abstract}
To deploy and operate a quantum network which utilizes existing
telecommunications infrastructure, it is necessary to be able to 
route entangled photons at high speeds, with
minimal loss and signal-band noise, and---most importantly---without disturbing
the photons' quantum state. 
Here we present a switch which fulfills these requirements and characterize its
performance at the single photon level; it exhibits a 200-ps switching
window, a 120:1 contrast ratio, 1.5 dB loss, and induces no
measurable degradation in the switched photons' entangled-state fidelity ($< 0.002$).  
Furthermore, because this type of switch couples the temporal and spatial degrees
of freedom, it provides an important new tool with which
to encode multiple-qubit states in a single photon.
As a proof-of-principle demonstration of this capability, 
we demultiplex a single quantum channel from
a dual-channel, time-division-multiplexed entangled photon stream,
effectively performing a controlled-bit-flip on a two-qubit
subspace of a five-qubit, two-photon state.
\end{abstract}

\maketitle

Switching technologies enable networked rather than point-to-point
communications.  Next-generation photonic quantum networks will
require switches that operate with low loss, low signal-band noise,
and \emph{without} disturbing the transmitted photons' spatial,
temporal, or polarization degrees of freedom \cite{mike_and_ike}.  Additionally,
the switch's operational wavelength must be compatible with
a low-loss, non-dispersive transmission medium, such as 
standard optical fiber's 1.3-$\mu$m zero-dispersion band \cite{nweke, oband}.
Unfortunately, no previously demonstrated technology
\cite{previous-first}--\cite{previous-last} is capable of
\emph{simultaneously} satisfying each of the above requirements:
waveguide electro-optic modulators (EOMs) \cite{eospace} and 
resonators \cite{waveguide_res1, waveguide_res2}
can operate at very high speeds (10 GHz) but completely destroy any
quantum information stored in the polarization degree of freedom;
micro-electromechanical switches \cite{mems1, mems2} do not degrade the photon's
quantum state, but operate at very low speeds ($<=250$ kHz);
polarization-independent EOMs \cite{eospace} 
can operate at moderate speeds ($\sim$100 MHz) but with
relatively high loss; and finally, traditional 1550-nm devices
based on nonlinear-optical fiber loops \cite{nolm1, nolm2} generate unacceptably high
levels of Raman-induced noise photons ($> 1$ in-band noise photon
per 100-ps switching window \cite{cband_noise}).

Although the requirements for ultrafast entangled-photon switching
are collectively daunting, they describe a device that is capable
of selectively coupling the spatial and temporal degrees of
photonic quantum information. In other words, a device that can
encode multiple-qubit quantum states onto a single photon, 
enabling quantum communication protocols that exploit
high-dimensional spatio-temporal encodings.  In this Letter
we describe the construction and characterization of
an all-optical switch which meets each of the aforementioned
requirements, and whose aggregate performance (in terms of
loss, speed, and in-band noise) exceeds that of all available
alternatives \cite{previous-first}--\cite{nolm2} by orders of magnitude
\cite{orders}.  Moreover, this switch design is scalable: by its extension
one can create devices that are capable of coupling
many temporal qubits and many spatial qubits.  As a proof-of-principle demonstration of
this capability, we utilize the
switch to perform a controlled-bit-flip operation on a two-qubit subspace
of a two-photon, five-qubit system, where a temporally encoded qubit is
used as the control and a spatially encoded qubit is used as the target.
This operation is used to demultiplex a single
quantum channel from a dual-channel, time-division-multiplexed
entangled photon stream encoded into the larger five-qubit space.  

\begin{figure}
\centering
\includegraphics[width=3in]{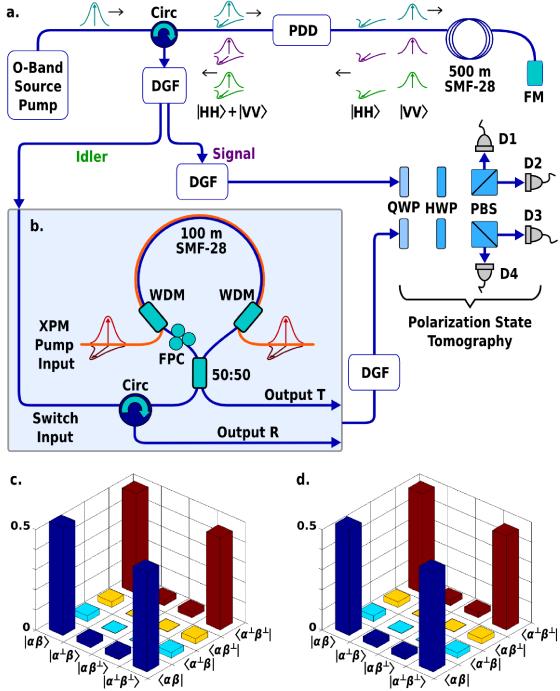}
\vspace{-1mm}
\caption{
(a) Entangled
photon-pair source and test apparatus for ultrafast switching. Nondegenerate
entangled photon pairs ($\lambda_\mathrm{signal} = 1303.5$ nm and
$\lambda_\mathrm{idler} = 1306.5$ nm) are
generated in 500 m of standard single-mode fiber (SMF-28). Signal and idler
photons are separated using a double-pass grating filter (DGF).
Circ: circulator, FM: Faraday mirror, FPC: fiber polarization controller, HWP:
half-wave plate, PBS: polarizing beam-splitter, PDD: polarization dependent
delay, QWP: quarter-wave plate, WDM: wavelength-division multiplexer. (b)
The single-photon switch. The length of the intra-loop SMF-28 is directly
proportional to the switching window. An $L$ = 100-m loop (shown) results
in a $\approx$200-ps switching window. 
(c) Reconstructed density matrix of the unswitched state for 
$L$ = 100-m (fidelity to a maximally entangled state, $F = 99.5\% \pm
0.2\%$).
(d) Density matrix of the switched state for $L$ = 100 m 
$(F = 99.4\% \pm 0.4\%)$.  Similar reconstructions for the 500-m switch (not
shown) yielded
$F = 99.5\% \pm 0.2\%$ (unswitched) and
$F = 99.2\% \pm 0.2\%$ (switched).
} \label{figure::switch}
\vspace{-3mm}
\end{figure}

In order to simultaneously achieve low loss and ultrafast switching, we
utilize an all-optical, fiber-based design in which bright 1550-nm pump
(C-band)
pulses control the trajectory of 1310-nm (O-band) single-photon signals
(see Fig.
\ref{figure::switch}(b)).  Physically, this switch exploits polarization-insensitive cross-phase
modulation (XPM) \cite{xpm} in a nonlinear-optical loop mirror (NOLM)
\cite{dual-lambda}, the
reflectivity of which is determined by the phase difference between the
clockwise and counter-clockwise propagating paths in a fiber Sagnac
interferometer (the ``loop'') \cite{loop-mirror}.  To actively control the state of
this switch, we initially configure an intra-loop fiber polarization controller 
such that the loop \emph{reflects} all incoming photons.  Multiplexing
a strong 1550-nm pump pulse into the clockwise or counter-clockwise loop
path then creates an XPM-induced phase shift on the respective clockwise or
counter-clockwise signal amplitude, with a $\pi$ phase shift causing the switch
to \emph{transmit all} incoming photons.  As XPM is inherently polarization
dependent, and polarization is often used to encode quantum information, it
is important that the pump pulse itself be effectively unpolarized.  
We accomplish this by temporally overlapping two orthogonally
polarized pump pulses, each with a slightly different wavelength
\cite{xpm}.

Note that the traditional NOLM-based C-band devices are unsuitible for single-photon
switching for two reasons:  Firstly, and most importantly, 
such switches generate very high levels of Raman-induced background
photons at signal wavelenths \cite{cband_noise}.  
These noise photons would swamp 
any single-photon signals, effectively ``washing out'' any two-photon quantum correlations.
Secondly, traditional NOLM-based devices utilize pump pulses which are
group-velocity matched to the signals being switched.  While this increases
the interaction time, the nonlinear character of
the XPM process limits the switching contrast in this type of operation.
Because the pump pulse can not be made perfectly square-shaped,
the center
of the signal pulse receives a stronger nonlinear phase shift than the pulse
wings, making it effectively impossible to choose a single pump power which
maximizes switching contrast over the entire signal pulse.


Our switch design is immune to each of these two fundamental problems and
allows for noiseless, high contrast switching operation. Firstly, the large
anti-Stokes detuning  ($\approx$35 THz) between the 1550-nm pump pulses
and the 1310-nm single-photon pulses reduces
contamination of the quantum channels by spontaneous Raman scattering of
the pump ($\approx 2 \times 10^{-7}$ background photons per ps of
signal pulse). 
Secondly, in standard single-mode fiber (the loop medium)
this detuning leads to a large group-velocity difference ($\approx$2 ps/m) between
the pump and signal pulses. This allows the switch to operate in a regime
where the pump pulse walks completely through the signal's temporal mode,
providing the type of uniform phase-shift which is essential for
high-contrast switching operation. The effective phase shift is therefore
determined by the total energy in a single pump pulse, regardless of that
pulse's temporal profile.  
The switching \emph{window}, $\tau$, is in turn
determined by the \emph{length of the
fiber} between the WDMs, $L$, multiplied by the speed at which the signal sweeps through the
pump (i.e., the group velocity difference between signal and pump).
For our case, $\tau = L \times 2$ ps/m.  The turn-on time of this switching window is
set by the temporal extent of the pump pulses (i.e., the time it
takes for the pump to physically enter and leave the fiber loop).  For 5-nm bandwidth,
transform-limited C-band pump
pulses, for example, the turn-on time can be as short as 1 ps.

Two key experimental technologies are
required to operate and
characterize this type of switch: a short-pulse 
dual-wavelength 1550-nm pump and a source of 1310-nm entangled
photons.
To create the dual-wavelength pump, two 5-ps duration pulses (1545-nm and 1555-nm
wavelengths) are spectrally carved with diffraction gratings from the output of a
50-MHz repetition rate mode-locked fiber laser (IMRA Femtolite Ultra, Model
BX-60), which are then multiplexed using a polarization beam combiner (PBC). 
The power necessary to produce a $\pi$
phase shift is obtained by amplifying the multiplexed pulses with 
a cascade of erbium-doped fiber amplifiers
(EDFAs). A long-pass filter with a 1543-nm edge is used after each EDFA to
ensure that the optical gain is confined to the pump pulses and that no
contaminating O-band photons are introduced by the pump preparation process.

The IMRA laser also provides an electrical clock signal for a 1310-nm entangled photon
source and an array of four single-photon detectors.  The entangled photon
source, described in detail in reference \cite{oband} and shown in Fig.
\ref{figure::switch}(a), utilizes spontaneous
four-wave-mixing in standard single-mode fiber to produce pairs of
polarization-entangled photons from
100-ps wide, 50-MHz repetition-rate pump pulses at 1305 nm.  After
switching, the photon pairs are measured with a correlated photon detection
system (Nucrypt LLC, Model CPDS-4) consisting of 
an array of four InGaAs/InP avalanche
photodiodes.

In order to test the switch's effectiveness for quantum communications, we
measured both active and passive switching of polarization-entangled photon
pairs.  Fig. \ref{figure::switch}(b) shows the switch as integrated
into the fiber-based source
of entangled photons referenced above.  To test multiple switching windows, loop
lengths of 500 m ($\approx$900-ps window) and 100 m ($\approx$180-ps window) were used. The
insertion loss introducecd by  these switches in the O-band quantum channel
was
measured to be 1.3 dB ($L$ = 100 m, port T), 1.7 dB ($L$ = 100 m, port
R), 1.7 dB ($L$ = 500 m, port T), and 2.1 dB ($L$ = 500 m, port R).  Because
all of these directly measured losses include the 0.4 dB or 0.8 dB loss from one or
two passes through an optical
circulator, the raw switching loss for either transmission through or
reflection from the switching loop is 
0.9 dB (1.3 dB) for the $L$ = 100-m (500-m) loop.

To set a performance
benchmark for the switch, unswitched (no pump) entangled
photons from port R were characterized using coincidence-basis
quantum-state tomography \cite{tomo1, tomo2}.  Both signal and idler photons were analyzed
using separate quarter waveplate (QWP), half waveplate (HWP), polarizing
beam splitter (PBS) combinations, which together perform arbitrary single
qubit measurements. The measured coincidence rates---after subtracting
accidental coincidences, a procedure which lowers statistical errors
\cite{accidentals}---for 36
combinations of analyzer settings \cite{detecting_entanglement} were subjected to a numerical
maximum likelihood optimization, which reconstructs the density matrix most
likely to have produced the measured results.
Figs. \ref{figure::switch}(c) and
\ref{figure::switch}(d), respectively, show reconstructed density matrices for 
passively switched (port R) and actively switched (port T) entangled
photons, after reflection or transmission through the $L$ = 100-m loop.
Similar reconstructions were performed for the $L$ = 500-m loop; 
in all four cases, the fidelity of the measured state
to a maximally-entangled state exceeded 99.0\%.  In addition, no measureable
state degradation resulted from active versus passive switching.

\begin{figure}
\centering
\includegraphics[width=3.25in]{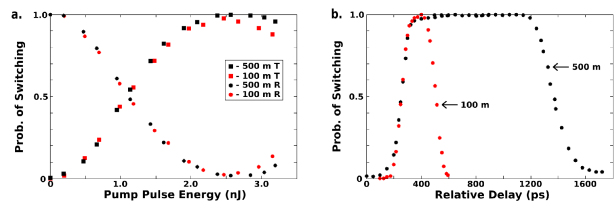}
\vspace{-1mm}
\caption{
(a) Single-photon switching
contrast.  Plot shows the probability of routing an incoming single photon
versus pump energy for $L = 500$ m and $L = 100$ m (detector dark counts
subtracted).  (b) Temporal extent of the switching window, as measured using
single photons.  Plot shows single-photon counts versus relative delay
between the single-photon and the pump pulses (detector dark counts
subtracted).
} \label{figure::characterization}
\vspace{-3mm}
\end{figure}

Another important metric for both quantum and classical
routers is the switching \emph{contrast}---or the ratio of
power directed to the desired output port divided by the
power directed to the complementary output port.  Fig.
\ref{figure::characterization}(a) shows the single-photon  
switching contrast as a function of the pump-pulse energy.
A contrast of 120:1 is achieved at a pump energy of 2.5 nJ for
$L$
= 500 m.  Although the switching contrast is expected to be independent of
$L$, the 100-m data does not appear to achieve full contrast---only 43:1.
This artifact is due to a long, low-power tail  
($\approx$370-ps total pulse width) in the
1305-nm pump pulses that drive the entangled photon source.  Although
the entangled photon
production rate is proportional to the pump-power \emph{squared}, this still
results in a longer than 200-ps pulse width for the entangled
photons used to test the switch.  As a result, the switching window is too
short to completely envelop the photon to be switched, resulting in an
artifically lower switching contrast for the $L$ = 100-m loop.  We expect the
true switching contrast to be the same in both cases ($\ge$ 120:1).

Closely related to contrast is the generated single-photon background,
from---for example---Raman scattering of the 1550-nm pump pulses.  We
measured the 
probability of generating a 1310-nm background photon count and found it to
be proportional to $L$
($\approx 4 \times 10^{-7} \mathrm{m}^{-1}$).  

In addition to its use as a single-photon router, the switch is a
spatio-temporal coupler, enabling the encoding and decoding of quantum
information into a temporally multimode Hilbert space, which is, in
principle, boundless.  The extent to which this Hilbert space can be
effectively accessed, however, is determined by the temporal switching
profiles of the devices described above.  To characterize the
shape and width of the switching window, we introduce a relative delay
between the signal and the pump pulses. Sweeping this delay while measuring
the switched single photons we map the switching window (shown in Fig.
\ref{figure::characterization}(b)).  
Note that the temporal extent of the photons being switched blurs the true switching
window.  To quantitatively estimate the extent of this blurring, we
directly measure the temporal shape of the 1305-nm
pump pulses used to create the test photons.  We perform this measurement by
constructing a third $L$ = 2-m switch with a switching window of
$\approx$10-ps (full width at half maximum).
Using the now-characterized 1305-nm classical pump pulses, we
re-measure the switching windows while 
applying a numerical-fit deconvolution to the results.  In this way we obtain
the instantaneous temporal widths of the 100-m and 500-m switching windows
to be 180 ps and 900 ps, respectively.

\begin{figure}
\centering
\includegraphics[width=3.25in]{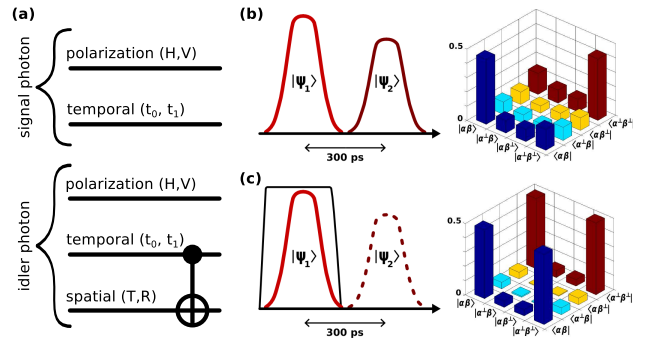}
\vspace{-1mm}
\caption{
(a) Diagram of the five degrees of freedom in a multiplexed entangled
photon stream, which can be demultiplexed by applying the controlled
switch operation (shown).  
(b) Arrangement of detected quantum
information channels for the five-qubit state
$\ket{\Phi} = c_1 \ket{\psi_1} \ket{t_0^s} \ket{t_0^i} \ket{T^i} + 
              c_2 \ket{\psi_2} \ket{t_1^s} \ket{t_1^i} \ket{T^i}$, 
where
$\ket{\psi_1} \equiv \frac{1}{\sqrt{2}} \left( \ket{H^s}\ket{H^i} + \ket{V^s}\ket{V^i} \right)$ 
and 
$\ket{\psi_2} \equiv \frac{1}{\sqrt{2}} \left( \ket{H^s}\ket{H^i} - \ket{V^s}\ket{V^i} \right)$.
Tracing over the temporal qubit and projecting into 
spatial mode $\ket{T^i}$, we reconstruct a highly mixed density
matrix with $F = 58.9\% \pm 0.5\%$.
(c) Arrangement of quantum information
channels after active switching to output T (i.e., demultiplexing), which should produce the
state:
$\ket{\Phi^\prime} = c_1 \ket{\psi_1} \ket{t_0^s} \ket{t_0^i} \ket{T^i} + 
              c_2 \ket{\psi_2} \ket{t_1^s} \ket{t_1^i} \ket{R^i}$.   
By projecting into spatial mode $\ket{T^i}$ we recover 
the maximally entangled state $\ket{\psi_1}$
($F = 98.6\% \pm 0.7\%$).
} \label{figure::mux}
\vspace{-3mm}
\end{figure}

The switch's ability to manipulate spatial and temporal quantum information has the
potential to enable new quantum communication protocols. As an example of
this functionality, we use the switch to demultiplex a single quantum
channel from a dual-channel entangled photon stream, on which we encode 
a five-qubit space (see Fig. \ref{figure::mux}(a)) defined by signal and idler polarization qubits
($\ket{H^{s,i}}, \ket{V^{s,i}}$),
signal and idler temporal qubits
($\ket{t_0^{s,i}}, \ket{t_1^{s,i}}$), and an idler spatial qubit
($\ket{T^{i}}, \ket{R^{i}}$---see Fig. \ref{figure::switch}(b)).
Using this encoding, we create the five-qubit hyper-entangled state 
$\ket{\Phi} = c_1 \ket{\psi_1} \ket{t_0^s} \ket{t_0^i} \ket{T^i} + 
              c_2 \ket{\psi_2} \ket{t_1^s} \ket{t_1^i} \ket{T^i}$, 
where
$\ket{\psi_1} \equiv \frac{1}{\sqrt{2}} \left( \ket{H^s}\ket{H^i} + \ket{V^s}\ket{V^i} \right)$, 
$\ket{\psi_2} \equiv \frac{1}{\sqrt{2}} \left( \ket{H^s}\ket{H^i} - \ket{V^s}\ket{V^i}
\right)$, and $c_1$ and $c_2$
are arbitrary coefficients.  Measuring $\ket{\Phi}$ using
polarization-basis tomography while \emph{tracing out} temporal
degrees of freedom and \emph{projecting} into the idler spatial mode $\ket{T^i}$ 
will yield a highly mixed state, exactly the result
one expects from a simultaneous measurement of multiple entangled
quantum channels.  A switch capable of implementing a controlled-NOT
operation which couples the spatial and temporal qubits (see Fig.
\ref{figure::mux}(a)),  
however, would transform $\ket{\Phi}$ into the state:
$\ket{\Phi^\prime} = c_1 \ket{\psi_1} \ket{t_0^s} \ket{t_0^i} \ket{T^i} + 
              c_2 \ket{\psi_2} \ket{t_1^s} \ket{t_1^i} \ket{R^i}$.   
              This demultiplexed state should exhibit maximal
entanglement when \emph{projected} into the spatial mode $\ket{T^i}$, because even after
tracing over the temporal degrees of freedom only the maximally entangled
polarization state $\ket{\psi_1}$ would be present.

To implement this proof-of-principle test of the switch's ability to couple
spatial and temporal degrees of freedom, 
we modify our
O-band entangled-photon source \cite{oband} by pumping it with a pair of 
pulses separated by 
$\Delta t \equiv t_1 - t_0 \approx 300 \mathrm{ps}$.
The polarizations of the leading and trailing pump pulses are chosen such that the
unnormalized pump state is
            $\sqrt{c_1} \left(\ket{H^p} +  \ket{V^p}\right) t_0 + 
             \sqrt{c_2} \left(\ket{H^p} + i\ket{V^p}\right) t_1$
which upon SFWM gives the output signal-idler state $\ket{\Phi}$.
For the demultiplexing test, we choose
$c_1/c_2 \approx 1.25$ and $\Delta t \approx300$ ps.  

Figure \ref{figure::mux}(b) shows
the experimentally measured density matrix for the multiplexed quantum
channels. As expected, the state is highly mixed; its fidelity to the
nearest maximally entangled state is only 58.9\%.  Utilizing the 100-m
switch we then
demultiplex (i.e., actively switch) only the first quantum channel 
($t = t_0$), creating the state $\ket{\Phi^\prime}$.  
As shown in Fig. \ref{figure::mux}(d), after demultiplexing 
we are able to recover the high fidelity (98.6\%) of the target state
to a maximally entangled state.  
Because the cross-Kerr phase shift has previously been shown to
maintain spatial and temporal coherence in NOLM switches \cite{nolm1,
nolm2}, we anticipate that this switch's 
cross-Kerr-based demultiplexing operation is in fact coherent and equivalent to the
controlled-NOT operation depicted in Fig. \ref{figure::mux}(a).
Moreover, unlike LOQC-based controlled-NOT gates, this switch is completely
deterministic and easily extensible, capable of independently tunable
couplings (e.g., 
controlled-$\pi/4$) to many temporal qubits encoded onto the same
photon  (by changing the control-pulse's
intensity as a function of time).  By cascading several switches, it is
also possible to couple to multiple spatial qubits.

In conclusion, we have demonstrated the first all-optical 
switch suitable for single-photon quantum communications. It achieves
low-loss ($< 1$ dB when used to switch between transmitted and reflected
modes, $< 1.7$ dB when combined with a circulator), 
high-isolation ($> 20$ dB), and high-speed ($< 200$ ps) performance without
a measureable disturbance to the quantum state of the routed single
photons. We demonstrate its ultrafast capability by demultiplexing a single
quantum channel from a time-division-multiplexed stream of entangled
photons.  Very few fundamental limitations apply to this type
of switch design.  With carefully designed fiber components, one has the
potential to dramatically reduce the switch's loss.  In principle the only
unavoidable switching losses are fiber transmission losses (0.15--0.2 dB/km) and circulator
insertion losses (waveguide-based circulators with a
0.05 dB insertion loss have been designed and simulated \cite{waveguide-circulators}).
Additionally, decreasing
$L$ to a few meters will reduce the switch's speed to
$\approx$10 ps
while simultaneously decreasing the background by an order of magnitude.  Even without
these improvements, however, this switch represents an important new tool
for manipulating spatianlly- and temporally-encoded quantum information.

This research was supported in part by the DARPA ZOE program (Grant No.
W31P4Q-09-1-0014) and the NSF IGERT Fellowship (Grant No. DGE-0801685).




\begin{thebibliography}{}
\vspace{-4mm}

\bibitem{mike_and_ike} M. A. Nielsen and I. L. Chuang, \emph{Quantum Computation
            and Quantum Information} (Cambridge Univ. Pr., 2000).
\bibitem{nweke} N. I. Nweke, \emph{et al.} 
            \emph{Appl.  Phys. Lett.} \textbf{87}, 174103 (2005).
\bibitem{oband} M. A. Hall, J. B. Altepeter, and P. Kumar, 
            \emph{Optics Express} \textbf{17}, 14558 (2009).
\bibitem{previous-first} M. A. Duguay and J. W. Hansen, 
            \emph{Appl. Phys. Lett.} \textbf{15}, 192 (1969).
\bibitem{previous2} N. J. Doran, and D. Wood, 
            \emph{Opt. Lett.} \textbf{13}, 56--58 (1988).
\bibitem{mems1} K. Hogari and T. Matsumoto,
            \emph{Appl. Opt.} \textbf{30}, 1253--1257 (1991).
\bibitem{nolm1} M. Eiselt, 
            \emph{Electron Lett.} \textbf{28}, 1505 (1992).
\bibitem{previous5} J. P. Sokoloff, P. R. Prucnal, I. Glesk, M. Kane, 
            \emph{IEEE Photon. Technol. Lett.} \textbf{5}, 787 (1993).
\bibitem{previous6} M. Asobe, I. Yokohama, H. Itoh, and T. Kaino, 
            \emph{Opt. Lett.} \textbf{22}, 274 (1997).
\bibitem{previous7} I. Yokohama \emph{et al.}, 
            \emph{J. Opt. Soc. Am. B} \textbf{14}, 3368 (1997).
\bibitem{previous8} G. S. Kanter, P. Kumar, K. R. Parameswaran, and M. M. Fejer, 
            \emph{IEEE Photon. Technol. Lett.} \textbf{13}, 341 (2001).
\bibitem{previous9} J. E. Sharping, M. Fiorentino, P. Kumar, and R. S. Windeler, 
            \emph{IEEE Photon. Technol. Lett.} \textbf{14}, 77 (2002).
\bibitem{previous10} V. Van, \emph{et al.} 
            \emph{IEEE Photon. Technol. Lett.} \textbf{14}, 74 (2002).
\bibitem{previous11} V. R. Almeida, \emph{et al.} 
            \emph{Opt.  Lett.} \textbf{29}, 2867 (2004).
\bibitem{previous-last} G. Bertocchi, \emph{et al.} 
            \emph{J. Phys. B} \textbf{39} 1011 (2006).
\bibitem{eospace} http://www.eospace.com
\bibitem{waveguide_res1} V. R. Almeida \emph{et al.}, 
            \emph{Opt. Lett.} \textbf{29}, 2867--2869 (2004).
\bibitem{waveguide_res2} P. Dong, S. F. Preble, and M. Lipson, 
            \emph{Opt. Express} \textbf{15}, 9600--9605 (2007).
\bibitem{mems2} C. Knoernschild, C. Kim, F. P. Lu, and J. Kim,
            \emph{Opt. Express} \textbf{17}, 7233--7244 (2009).
\bibitem{nolm2} K. Uchiyama \emph{et al.},
            \emph{J. Lightw. Tech.} \textbf{15}, 194--201 (1997).
\bibitem{cband_noise} Q. Lin, F. Yaman, and G. P. Agrawal, 
            \emph{Phys. Rev. A} \textbf{75}, 023803 (2007).
\bibitem{orders} For simplicitly, we define ``aggregate performance'', in
dB units, as: (relative loss) + (relative speed) + (relative production rate of
in-band noise photons).
\bibitem{xpm} H. B\"{u}low and G. Veith, 
            \emph{Elect. Lett.} \textbf{29}, 588--589 (1993).
Note that if one of the two pump colors leads the other by a
time $\delta$, then the leading and trailing $\delta$-length segments of the
switching window will not be polarization-independent.
\bibitem{dual-lambda} K. J. Blow, N. J. Doran, B. K. Nayar, and B. P. Nelson, 
            \emph{Opt. Lett.} \textbf{15}, 248--250 (1990).
\bibitem{loop-mirror} D. Mortimer, 
            \emph{J. Lightw. Tech.} \textbf{6}, 1217--12124 (1989).
\bibitem{tomo1} D. F. V. James, P. G. Kwiat, W. J. Munro, and A. G. White,
            \emph{Phys. Rev. A} \textbf{64}, 052312 (2001).
\bibitem{tomo2} J. B. Altepeter, E. R. Jeffrey, and P. G. Kwiat, 
            \emph{Advances in AMO Physics, Vol. 52}, Ch. 3 (Elsevier, 2006).
\bibitem{accidentals} 
Increasing this source's pair production rate (PPR)
and then subtracting accidental coincidences increases measurement
precision while accurately predicting the low-PPR, non-accidental
subtracted result \cite{oband}.  Without this correction, high-PPR,
dark-count-only subtracted fidelities for the data shown in Figs. 1 and 3 are between
80--95\%.
\bibitem{detecting_entanglement} J. B. Altepeter, \emph{et al.}
            \emph{Phys. Rev. Lett.} \textbf{95}, 033601 (2005).
\bibitem{waveguide-circulators} R. Takei and T. Mizumoto,
\emph{Jpn. J. Appl. Phys.} \textbf{49}, 052203 (2010).

\end{thebibliography}
\end{document}